\input{aipcheck}

\documentclass[
    ,final            % use final for the camera ready runs
  ]
  {aipproc}

\layoutstyle{6x9}
\usepackage{bm}

\begin{document}

\title{Neutrino Magnetic Moment}

\classification{14.60.Pq, 13.40.Em, 26.65.+t, 96.15.Gh}
\keywords      {Neutrino magnetic moment, spin-flavor  precession, solar 
antineutrinos}

\author{A.B. Balantekin}{
  address={University of Wisconsin, Physics Department, Madison, 
WI 53706 USA\footnote{E-Mail: baha@physics.wisc.edu}}
}

\begin{abstract}
Current experimental and observational limits on the neutrino magnetic 
moment are reviewed. Implications of the recent results from the solar 
and reactor neutrino experiments for the value of the neutrino magnetic 
moment are discussed. It is shown that spin-flavor precession in the Sun 
is suppressed. 
\end{abstract}

\maketitle

\section{Introduction}

A minimal extension of the Standard Model (with non-zero neutrino masses) 
yields a neutrino magnetic moment of \cite{Marciano:1977wx}
\begin{equation}
\mu_{\nu} = \frac{3eG_F m_{\nu}}{8 \pi^2 \sqrt{2}} = 
\frac{3G_F m_e m_{\nu}}{4 \pi^2 \sqrt{2}} \mu_B
\label{a1}
\end{equation}
where $\mu_B = e/2m_e$ is the Bohr magneton. Note that the neutrino 
magnetic moment is proportional to the neutrino mass as required by the 
symmetry principles. Since the recent solar, atmospheric and reactor 
neutrino experiments indicate the existence of non-zero neutrino masses, 
we also know that neutrino has a magnetic moment. Using the neutrino 
parameters deduced from analyses of those experiments 
\cite{Balantekin:2003dc} we get $\mu_{\nu} \geq ( 4 \times 10^{-20}) 
\mu_B$. Larger values of 
magnetic moments are possible in extensions of the Standard Model, as 
indicated by the inequality sign in this value. If the 
magnetic moment is generated by physics at scale $\Lambda$ we can write 
\begin{equation}
\mu_{\nu} \sim \frac{e {\cal G}}{\Lambda} , 
\label{a2}
\end{equation}
where ${\cal G}$ represents the combination of the coupling constants 
and appropriate $ 2\pi$ factors.  
If we remove the external photon from the diagrams leading to Eq. 
(\ref{a2}) we get a contribution to the mass of the order
\begin{equation}
\delta m_{\nu} \sim {\cal G} \Lambda . 
\label{a3}
\end{equation}
These equations imply that 
\begin{equation}
\delta m_{\nu} \sim \frac{\Lambda^2}{m_e}  \left( \frac{\mu_{\nu}}{\mu_B} 
\right). 
\label{a4}
\end{equation}
If one assumes that the scale $\Lambda$ is not significantly higher than the 
electroweak scale, current neutrino mass limits imply 
$|\mu_{\nu}| \leq 10^{-14} \mu_B$ for Dirac neutrinos \cite{Bell:2005kz}. 
It is however possible to introduce models where the magnetic moment and mass 
do not come from the same number of loops, and relax this bound (see e.g. 
Ref. \cite{Georgi:1990se}). 

It is well-established that the neutrinos mix and 
the discussion above illustrates that magnetic moment is properly defined 
in the mass basis \cite{Beacom:1999wx}. In this basis Dirac neutrinos can have 
both diagonal and off-diagonal moments, whereas Majorana neutrinos can only 
have transition moments. More specifically, if the magnetic moment operator 
is designated by $\bm{\mu}$, then $\bm{\mu} = {\bm{\mu}}^{\dagger}$ for 
Dirac neutrinos, and $\bm{\mu}^T = -  
\bm{\mu}$ for Majorana neutrinos. 

\section{Limits on the Neutrino Magnetic Moment}

There are a number of possible 
physical processes involving a neutrino with a magnetic 
moment. Among these are the $\nu - e$ scattering, spin-flavor precession 
in an external magnetic field, plasmon decay, and the neutrino decay. 
For the first process, using the magnetic moment operator $\bm{\mu}$, the 
total cross section at an experiment  where the final neutrino is not observed 
can be written in the Born approximation as 
\begin{equation}
\sigma \sim \sum_i | \langle \nu_i | \bm{\mu} | \nu_e \rangle |^2 ,
\label{a5}
\end{equation}
where $| \nu_i \rangle, i=1,2,3$ represent the mass eigenstates and we 
assumed that electron neutrinos are used. Since 
the neutrino mixing matrix in  $| \nu_e \rangle = \sum_i U_{ei} 
| \nu_i \rangle$ is unitary, Eq. (\ref{a5}) takes the form 
\begin{equation}
\sigma \sim \langle \nu_e | \bm{\mu}^{\dagger} \bm{\mu} | \nu_e \rangle .
\label{a6}
\end{equation} 
Detecting a neutrino magnetic moment then implies detecting them in mass 
eigenstates. Consequently the measured magnetic moment of the neutrino, 
in principle, depends on the distance from its source \cite{Beacom:1999wx}:
\begin{equation}
\mu_e^2 = \sum_i \mid \sum_j U_{ej} \mu_{ij} \exp ( -iE_jL) \mid^2 .
\label{a7}
\end{equation}

The differential scattering 
cross section for electron neutrinos or antineutrinos 
on electrons is given by \cite{Vogel:1989iv}
\begin{eqnarray}
\frac{d\sigma}{dT} &=& \frac{G_F^2m_e}{2\pi} 
\left[(g_V + g_A)^2 + (g_V - g_A)^2 \left(1-\frac{T}{E_\nu}\right)^2 + 
(g_A^2-g_V^2) \frac{m_eT}{E_\nu^2}\right] \nonumber \\
&+& \frac{\pi \alpha^2 \mu_{\nu}^2}{m_e^2} \left[ \frac{1}{T} - 
\frac{1}{E_{\nu}} \right], 
\label{d1}
\end{eqnarray}
where T is the electron recoil kinetic energy, $g_V = 2 \sin^2 \theta_W 
+ 1/2$, $g_A = +1/2 (-1/2)$ for electron neutrinos (antineutrinos), and the 
neutrino magnetic moment is expressed in units of $\mu_B$. The first line 
in Eq. (\ref{d1}) is the standard electroweak contribution and the second line 
represents the contribution of the neutrino magnetic moment. Clearly the 
magnetic moment contribution is dominant at low recoil energies. The 
magnetic moment cross section will exceed the standard electroweak 
cross-section for recoil energies
\begin{equation}
\frac{T}{m_e} < \frac{\pi^2 \alpha^2}{(G_F m_e^2)^2} \mu_{\nu}^2, 
\label{d2}
\end{equation}  
i.e. the lower the smallest measurable recoil energy is, the smaller 
values of the magnetic moment can be probed. To perform such an experiment 
either solar or reactor neutrinos have been used. 
SuperKamiokande collaboration 
looked for distortions in the energy spectrum of solar neutrinos scattered 
off the electrons in their detector. No clear signal was observed. Combined 
with the other solar neutrino and KamLAND experiments a limit of 
$\mu_{\nu} \leq 1.1 \times 10^{-10}$ $\mu_{B}$ at 90\% C.L. was 
obtained \cite{Liu:2004ny}. The MUNU collaboration, using reactor neutrinos,  
recently obtained a slightly better limit of 
$\mu_{\nu} \leq 9 \times 10^{-11}$ $\mu_{B}$ at 90\% C.L. 
\cite{Daraktchieva:2005kn}. Another possibility for doing such experiments 
is to utilize low-energy beta beams \cite{Volpe:2003fi}. A detailed study 
of neutrino-electron scattering using low-energy beta-beams in general 
is given in Ref. \cite{Balantekin:2005md} and limits on the neutrino 
magnetic moment in particular are given in Ref. \cite{McLaughlin:2003yg}. 
The latter work finds that a tritium source may yield a better bound. 

Neutrinos change helicity in magnetic moment scattering. This fact has 
been used to put limits from astrophysics and cosmology on neutrino magnetic 
moment. If $\mu_{\nu}$ is sufficiently large, then the proto-neutron star 
formed in a core-collapse supernova can cool faster since the right-handed 
components are sterile. It was found that $\mu_{\nu} \ge 10^{-12} \mu_B$ 
would be inconsistent with the observed cooling time of SN1987a 
\cite{Lattimer:1988mf}. In the Early Universe the existence of 
right-handed Dirac neutrinos that may be produced in magnetic scattering 
increase the number 
of effective degrees of freedom altering neutrino counting through the 
big-bang nucleosynthesis yields \cite{Morgan:1981zy}. (Similar limits do not 
apply to Majorana neutrinos since antineutrino states are already counted). 

The tightest astrophysical bound on neutrino magnetic moment comes from the 
red-giant stars. A large enough magnetic moment implies enhanced plasmon 
decay rate, $\gamma^* \rightarrow \nu \nu$, inside the star. Since the 
neutrinos freely escape the stellar environment this process in turn cools 
a red giant star faster, delaying helium ignition. Existing observations 
of globular cluster stars lack any evidence of this effect, yielding a 
limit of $\mu_{\nu} \le 3 \times 10^{-12} \mu_B$ 
\cite{Raffelt:1990pj}. 

The discussion above shows the neutrino magnetic moment is presently known 
to be in the range
\begin{equation}
(9 \times 10^{-11}) \mu_B  \ge \mu_{\nu} \ge (4 \times 10^{-20}) \mu_B .
\end{equation}
The large width of this range represents possible physics beyond the 
standard model which can be explored using the neutrino magnetic moment 
measurements. 

\section{Implications of Neutrino Spin-Flavor Precession}

If neutrinos have magnetic moments, large magnetic fields that exist in 
astrophysical environments may give rise to an additional 
spin-flavor 
precession coupled to the usual matter-enhanced neutrino oscillations 
\cite{Lim:1987tk,Balantekin:1990jg}. Spin-flavor precession changes the 
helicity of the neutrinos, and if the neutrinos are of Majorana type, this 
yields a solar antineutrino flux 
\cite{Raghavan:1991em,Akhmedov:1991uk,Balantekin:1992dv}. 
Since the electron antineutrino yields a very distinctive two-neutron signal 
on charged-current deuteron break-up, Sudbury Neutrino Observatory (SNO) 
measurements were able to 
put a limit of $\Phi_{\bar{\nu_e}} \le 3.4 \times 10^4 {\rm cm}^{-2} 
{\rm s}^{-1}$ at 
90\% C.L. \cite{Aharmim:2004uf}. This corresponds to less than 0.8\% of the 
standard solar model $^8B$ flux. The KamLAND, experiment, being directly 
sensitive to antineutrino scattering in their scintillator liquid, 
provides a slightly better bound of 
$\Phi_{\bar{\nu_e}} \le 3.7 \times 10^2 {\rm cm}^{-2} 
{\rm s}^{-1}$ at 
90\% C.L. \cite{Eguchi:2003gg}. 
This is less than $2.8{\times}10^{-4}$ of the Standard Solar Model $^8$B 
$\nu_e$ flux.

A complete analysis of the spin-flavor precession scenario in the Sun 
requires detailed knowledge of the solar magnetic fields. Unfortunately  
information about solar magnetic fields is rather incomplete. If the 
magnetic field is greater than $10^8$ G, magnetic pressure becomes the 
same order of magnitude as the matter pressure obviating the Standard Solar 
Model. For the neutrino masses and the mixing angles deduced from the solar 
and reactor neutrino experiments, 
both the spin-flavor and the MSW resonances are very 
close together in the inner radiative zone. It was shown that 
magnetic fields greater than 
$\sim 10^7$ G, localized at about 0.2 R$_{\odot}$, would cause the sound 
speed profile to be at variance with the helioseismic observations 
\cite{Couvidat:2003ba}. 

The MSW resonance takes place in the Sun where the condition 
\begin{equation}
\label{p1}
\sqrt{2} G_F N_e = \frac{\delta m^2}{2E_{\nu}} \cos 2\theta
\end{equation}
is satisfied. The spin-flavor precession resonance takes place before 
the solar neutrinos reach the MSW resonance point. It is where 
\begin{equation}
\label{p2}
\frac{G_F}{\sqrt{2}} ( 2 N_e - N_n) = \frac{\delta m^2}{2E_{\nu}}
\cos 2\theta
\end{equation}
for the Dirac neutrinos and where 
\begin{equation}
\label{p3}
\sqrt{2} G_F ( N_e - N_n) = \frac{\delta m^2}{2E_{\nu}} \cos 2\theta
\end{equation}
for the Majorana neutrinos. In these equations $N_e$ and $N_n$ are the 
electron and neutron densities, respectively. For the observed neutrino 
parameters these resonances significantly overlap, hence previous 
approaches treating them as isolated resonances \cite{Balantekin:1992dv}
are not applicable \cite{Balantekin:2004tk,Friedland:2005xh}. However, 
there is a limit which may provide an analytical insight into the problem of 
overlapping resonances. In the limit $N_n=0$, clearly the resonances of Eqs. 
(\ref{p1}), (\ref{p2}), and (\ref{p3}) are at the same location. 
The electron neutrino survival probability is \cite{Balantekin:1998yb}
\begin{equation}
  \label{p4}
   P(\nu_e \rightarrow \nu_e) = \frac{1}{2} - \frac{1}{2}
\cos{2\theta}  \left( 1
- 2 P_{\rm hop} \right),
\end{equation}
where $P_{\rm hop}$ is the hopping probability between matter eigenstates. 
It can be shown that \cite{Balantekin:2004tk}, in the limit $N_n=0$, the 
hopping probability is given by
\begin{eqnarray}
\label{p5}
P_{hop} (\mu B \neq 0) = P_{hop} (\mu B = 0)
\times \nonumber \\
\exp  \left\{
\frac{i}{\pi}
\int^{r_0^*}_{r_0} dr \frac{\delta m^2}{2 E} \left[ \frac{(\mu B)^2}{
\sqrt{ \zeta^2(r) - 2\zeta(r)\cos{2\theta_v} +1 }} \right] \right\}, 
\end{eqnarray}
where we used the semiclassical treatment of the matter-enhanced 
neutrino oscillations to calculate the hopping probability 
\cite{Balantekin:1996ag}. 
In Eq. (\ref{p5}), $\zeta(r) = \frac{2\sqrt{2} G_F N_e(r)}{\delta m^2/E}$, 
and  ${r_0^*}$ and ${r_0}$ are the turning points (zeros) of the
integrand. 
Matter-enhanced neutrino oscillations in the Sun are primarily adiabatic 
\cite{Balantekin:2003dc}, hence the hopping probability is very small 
to begin with. Eq. (\ref{p5}) implies that the reduction factor 
of the hopping probability,  
\begin{equation}
\label{p6}
\exp \left[ - \frac{\pi}{\alpha} \frac{(\mu B)^2 2 E}{\delta m^2}
\right], 
\end{equation}
is also very small. 
For a $10^5$ G magnetic field, a magnetic moment of $10^{-12} \mu_B$,
and $E \sim 10 $  MeV, this hopping reduction factor is $ \sim 10^{-3}$.
An exact numerical calculation, using the neutrino mixing parameters 
$\delta m^2 = 8 \times 10^{-5}$ eV$^2$, $\tan^2 \theta = 0.4$, and 
relatively large values of $\mu_{\nu} = 10^{-11} \mu_B$ and $B = 10^5$ G, 
finds that the electron neutrino survival probabilities 
calculated with the MSW resonance only ($B=0$) and calculated with both 
resonances ($B \neq 0$) differ by less than $10^{-5}$ 
\cite{Balantekin:2004tk}.  
For reasonable values of the solar magnetic fields the effect of the 
neutrino magnetic moment on 
the solar neutrino flux is minuscule. 
It should be noted however that large fluctuations of the magnetic fields 
could impact spin-flavor precession \cite{Loreti:1994ry}. 

\begin{theacknowledgments}
This work was supported in part by the U.S. National Science Foundation 
Grant No. PHY-0244384, and in part by the University of Wisconsin Research 
Committee with funds granted by the Wisconsin Alumni Research Foundation.
\end{theacknowledgments}

\end{document}